\documentclass[12pt]{article}

\usepackage{scicite}
\usepackage{graphicx,color,epsfig,times,hyperref}

\newenvironment{sciabstract}{%
\begin{quote} \bf}
{\end{quote}}

\newcounter{lastnote}
\newenvironment{scilastnote}{%
\setcounter{lastnote}{\value{enumiv}}%
\addtocounter{lastnote}{+1}%
\begin{list}%
{\arabic{lastnote}.}
{\setlength{\leftmargin}{.22in}}
{\setlength{\labelsep}{.5em}}}
{\end{list}}

\title{Surface Crystallization in a Liquid AuSi Alloy.}

\author
{Oleg~G.~Shpyrko$^{1,2\ast}$, Reinhard Streitel$^{1}$,\\
Venkatachalapathy S. K. Balagurusamy$^{1}$,
Alexei~Y.~Grigoriev$^{1}$,\\ Moshe~Deutsch$^{3}$,
Benjamin~M.~Ocko$^{4}$, Mati Meron$^{5}$,\\ Binhua
Lin$^{5}$, Peter~S.~Pershan$^{1}$\\
\\
\normalsize{$^{1}$Department of Physics and Division of
Engineering and Applied Sciences,}\\
\normalsize{Harvard University, Cambridge, MA 02138, USA}\\
\normalsize{$^{2}$Center for Nanoscale Materials, Argonne National Laboratory, Argonne, IL 60439, USA}\\
\normalsize{$^{3}$Department of Physics, Bar-Ilan University, Ramat-Gan 52900, Israel}\\
\normalsize{$^{4}$Condensed Matter Physics and Materials Science
Department,}\\
\normalsize{Brookhaven National Laboratory, Upton NY 11973, USA}\\
\normalsize{$^{5}$Center for Advanced Radiation Sources, University of Chicago, Chicago, IL 60637, USA}\\
\normalsize{$^\ast$To whom correspondence should be addressed.
E-mail:  oshpyrko@anl.gov.}\\
\normalsize{4 April 2006; accepted 25 May 2006;
\textcolor[rgb]{0.00,0.00,1.00}{
\href{http://dx.doi.org/10.1126/science.1128314}{doi:10.1126/science.1128314}}}}
\date{}

\begin{document}
\baselineskip24pt
\maketitle
\begin{sciabstract}
X-ray measurements reveal a crystalline monolayer at the surface
of the eutectic liquid Au$_{82}$Si$_{18}$, at temperatures above
the alloy's melting point. Surface-induced atomic layering, the
hallmark of liquid metals, is also found below the crystalline
monolayer. The layering depth, however, is threefold greater than
that of all liquid metals studied to date. The crystallinity of
the surface monolayer is notable, considering that AuSi does not
form stable bulk crystalline phases at any concentration and
temperature and that no crystalline surface phase has been
detected thus far in any pure liquid metal or nondilute alloy.
These results are discussed in relation to recently suggested
models of amorphous alloys.
\end{sciabstract}

Surface melting - the coexistence of a liquid surface layer with
the bulk crystal at temperatures below the bulk melting point
$T_m$ - has been observed in a wide range of
materials\cite{Frenken85,Dash85} and occurs because the entropy of
molecules at the free surface is greater than that in the bulk due
to the reduced number of their near-neighbors. The opposite
effect, surface freezing, where a crystalline surface layer
coexists with its molten bulk, is much rarer. Surface freezing has
been observed, however, in complex liquids composed of
high-anisotropy molecules, such as molten unary or binary alkanes
and their derivatives\cite{Wu93}, and in liquid
crystals\cite{Dogic03}. Theory assigns the occurrence of this
effect to the highly anisotropic shape of the molecules and to
their lengths being greater than the interfacial
width\cite{Rabin}.

Freezing of the surface-segregated component into a
two-dimensional (2D) solid layer has also been reported recently
in the very dilute binary metallic alloys
Ga$_{99.948}$Pb$_{0.052}$\cite{Yang99} and
Ga$_{99.986}$Tl$_{0.014}$\cite{Yang03}. A different ordering
effect, surface-induced layering consisting of stratified layers
(Fig. 1) near the vapor interface\cite{Magnussen95,Regan95}, has
been observed in all liquid metals and alloys studied to date. The
decay of the layering order with depth is exponential and has a
range equal to the bulk liquid correlation length (two to three
atomic diameters). No surface-parallel ordering was found within
these layers in any elemental liquid metal. Similar layering,
along with epitaxially induced surface-parallel order, has also
been observed in both metallic and nonmetallic liquids near
solid/liquid interfaces\cite{Donnelly02,Oh05,Huisman97}.

We present evidence for surface crystallization and enhanced
surface layering in the liquid Au$_{82}$Si$_{18}$ eutectic alloy
of a type unlike that previously reported for any liquid metal or
alloy. A surface monolayer that exhibits lateral long-range
crystalline order was found above the eutectic temperature
$T_e=$359~$^{\circ}$C. Beneath this monolayer, seven to eight
layers occur that are liquid in the lateral direction but well
defined in the normal direction. The crystalline surface monolayer
and the enhancement of the surface-induced layering range beyond
the two to three layers observed in all other liquid metals
studied to date clearly have a common origin. The surface-frozen
monolayer undergoes a first-order transition into a different
surface phase 12~$^\circ$C above $T_e$.

These unusual surface structures probably result from the equally
unusual bonding properties of metastable amorphous bulk AuSi.
AuSi, the first metallic alloy found to exhibit a glassy solid
phase\cite{Duwez60}, remains one of the most puzzling amorphous
solids. Silicon-rich amorphous AuSi is a semiconductor, has a low
packing density, and has a low atomic coordination number (4 to
5). Its main structural motif is a continuous random network of
covalently bonded Si atoms. The Au-rich alloy, however, is a
metallic glass, almost as dense as a face-centered cubic lattice,
has a high atomic coordination number (8 to 9), and has a random
hard-sphere packing \cite{Weis92}. Such random packing in
amorphous metals was recently shown to consist of interpenetrating
clusters, the outer atoms of which are shared by adjacent
clusters\cite{Miracle04,Sheng06}. Because Si has a lower surface
tension than Au, the surface of liquid AuSi alloy is Si-rich, so
atomic packing and bonding at the surface might be expected to be
more like that of the covalently bonded Si-rich alloys than the
metallic Au-rich bulk.

The existence of a very deep eutectic (for Au$_{82}$Si$_{18}$) at
$T_e\approx$359~$^\circ$C, much below $T_m=$1063~$^\circ$C of Au
and $T_m=$1412~$^\circ$C of Si (Fig. 2A, inset), arises from the
bonding effects discussed above. Below $T_e$, AuSi phase-separates
in thermodynamic equilibrium into crystalline Au and Si, with no
mutual solid solubility and no stable crystalline intermetallic
compounds, whereas metastable amorphous AuSi phases can be
achieved by rapid quenching, sputtering and other
techniques\cite{Weis92}.

X-ray reflectivity off a liquid surface, $R(q_z)$, is measured as
a function of the grazing angle of incidence $\alpha$. Here
$q_z=(4\pi/\lambda)\sin \alpha$ is the surface-normal wavevector
transfer and $\lambda$ is the x-rays' wavelength. The ratio
$R/R_F$, where $R_F$ is the theoretical reflectivity off of an
ideally flat and abrupt liquid-vapor interface, depends on the
surface-normal electron density profile $\rho(z)$. A layered
interface produces a Bragg-like peak in $R(q_z)/R_F(q_z)$ because
of the constructive interference of the rays reflected from the
periodically ordered surface layers\cite{Magnussen95}. The larger
the number of layers, the higher is the layering peak. Fig. 2A
demonstrates that the AuSi layering peak at 370 $^\circ$C is at
least one order of magnitude higher than for the standard layering
profile observed in all elemental liquids measured to
date\cite{AuSn}. This result implies that there are more than the
two to three layers found in all previously measured liquid
metals.

Indeed, a theoretical model fit to the measured values of
$R(q_z)/R_F(q_z)$ (Fig. 2A, red line) yields the $\rho(z)$ curve
shown in Fig. 2B. Although the finer details of this $\rho(z)$
curve may not be unique, two features were independent of the
model used: the 2.5-nm-thick layering range (seven to eight
well-defined atomic layers) and the Si enrichment of the top
layer, indicated by a $\rho$-value less than that of
$\rho_{bulk}$, which corresponds to $\sim70$ atomic {\%} Si. This
value agrees well with the 67 atomic {\%} calculated from the
Gibbs adsorption rule for an ideal binary
solution\cite{Shpyrko05}.

No variation was found in the measured $R(q_z)$ from $T_e=$
359~$^\circ$C up to 371~$^\circ$C. At 371~$^\circ$C,
$R(q_z)/R_F(q_z)$ changed abruptly (Fig. 2A, curve with black
squares). By monitoring $R(q_z)$ at a fixed $q_z$ while varying
the temperature (Fig. 2B, inset), we found the surface phase
transition at 371~$^\circ$C to be reproducible and to exhibit no
hysteresis ($<$0.1$^\circ$C). The narrow width of 0.17~$^\circ$C
suggests a first-order phase transition.

The surface-parallel structure was explored by grazing incidence
x-ray diffraction (GIXD). X-rays impinging on the liquid AuSi
surface well below the critical angle penetrated the surface only
evanescently, to a depth of $\sim1.4$~nm \cite{GIXD}, and produced
a diffraction pattern for only the top $\sim5$ atomic surface
layers. The GIXD pattern measured for 359~$^{\circ}\textrm{C} \leq
T \leq$371~$^{\circ}$C (Fig. 3) showed sharp diffraction peaks
indicative of long-range lateral ordering. A broad peak,
characteristic of a liquid, was also observed. The GIXD pattern
was indexed in a 2D rectangular lattice of dimensions
\textbf{a}=7.386~{\AA} and \textbf{b}=9.386~{\AA}. A full-pattern
refinement yielded the Au$_4$Si$_8$ structure shown in the right
inset of Fig. 3. The high-T surface phase, which forms at
371~$^\circ$C and exists up to at least 410~$^\circ$C, also
exhibits a set of sharp GIXD peaks but at different $q_{xy}$
positions\cite{Future} than those of the 359~$^{\circ}\textrm{C} <
T < $371~$^{\circ}$C surface phase.

The GIXD peak intensities in the low-T phase were not affected by
sample rotation around the surface-normal axis, indicating that
the diffracting monolayer consists of a fine powder of randomly
oriented crystallites. Debye-Scherrer analysis of the line shapes,
measured with a high-resolution analyzer crystal, yields a typical
crystallite size of $\sim2$ to $10~\mu$m. The measured Bragg rods
(see the supporting online material) are surface-normal,
indicating a quasi-2D crystalline structure. The
$\sim1.5$~{\AA}$^{-1}$, width of the rod's $q_z$ intensity
distribution implies a crystalline layer thickness of
$d=\pi/(1.5\textrm{\AA}^{-1})$, where $d$ is approximately equal
to 2 $~{\AA}$, that is, a monolayer. This result agrees with $d$
value of $\approx 2.5$~{\AA} that is estimated from the ratio of
the integrated intensities of the GIXD peaks and the broad liquid
peak.

Notably, despite the crystalline order, the capillary surface
dynamics are still liquid-like; the diffuse scattering measured
away from the specular peak exhibits a power-law behavior
characteristic of the height-height correlations in liquid
surfaces\cite{Sinha88} (Fig. 3, left inset). The line shape of the
diffuse scan (solid line in left inset of Fig. 3) is well
reproduced by the capillary wave theory profiles if we use
$\gamma_{AuSi}= 780$~mN/m convolved with the experimental
resolution function. The compliance with capillary wave theory
also indicates that in spite of the larger layering depth, the
subsurface ordered layers are laterally liquid, as is the case for
all previously measured liquid metals.

 The crystalline AuSi monolayer's structure does
not resemble those of pure Au, pure Si or any of the reported
metastable bulk intermetalics. However, crystalline phases with
unit cell dimensions \textbf{a}=7.44~{\AA} and
\textbf{b}=9.33~{\AA} similar to those observed in our study, were
reported in thin Au films deposited on a Si(111)
surface\cite{Green76}. Some of these phases were thought to be
surface phases, and some may exist also in the bulk. No reliable
thickness of these phases could be obtained by the low-energy
electron diffraction and Auger electron spectroscopy techniques
that were used. A clear understanding of the formation mechanism,
stability, and surface specificity of the crystalline phases in
this system has not yet emerged\cite{Chang04}. However, the strong
Si-Si bonding\cite{Chen67} was suggested to play a crucial role,
as well as the quasi-2D atomic coordination near the surface
\cite{Molodtsov91}. In our case, the Gibbs adsorption surface
enrichment by Si atoms is likely to facilitate the chemical Au-Si
bonds that stabilize the crystalline surface layer but which
evolve to metallic-like bonding away from the
surface\cite{Filonenko69}.

The Si packing in the crystalline surface monolayer (Fig. 3, right
inset) resembles the covalently bonded, network-forming Si chains
suggested to stabilize the amorphous structure  of Si-rich AuSi
alloys\cite{Miracle04,Sheng06}. A comparison of the unit cell
structure presented in Fig. 3 with fig. 1A of
Miracle\cite{Miracle04} indicates  that the unit cell packing here
is close to the interpenetrating cluster structure recently
suggested as the main structural motif of bulk amorphous solids.

The formation of a surface-frozen phase is typically marked by a
change in the slope of the surface tension versus temperature
curve, $\gamma(T)$, from $d\gamma/dT > 0$ for a crystalline
surface to $d\gamma/dT < 0$ for a liquid surface\cite{Wu93}.
Indeed, a positive slope is reported for AuSi above
melting\cite{Bischoff00}, in line with the crystalline phase
reported here. The change to a negative slope, reported to occur
at $T \approx$ 800 to 900~$^\circ$C\cite{Bischoff00}, should
indicate the melting of the ordered surface structure. Similar
positive $\gamma(T)$ slopes above melting have been recently
reported for AgSn, AgBi, AgIn, InCu, CuSn, MnSn and AuZn binary
alloys, implying that the formation of surface-frozen phases may
be not an entirely uncommon phenomenon in multi-component liquid
alloys.

Beyond its importance for understanding the physics underlying
amorphous metallic alloys, AuSi is also of high technological
importance, because Au is widely used in interconnecting
integrated circuits on Si substrates. AuSi also has important
nanoscale applications such as the self-assembly of Si
nanowires\cite{Hu99} and low-temperature bonding in micro- and
nanoelectromechanical devices\cite{Cheng00}. Surface phases are of
particular importance for nanotechnology, because properties of
objects at the nanometer scale are expected to be dominated by
surfaces and interfaces. The discovery of previously unidentified
structures bridging the gap between 2D and 3D phases is expected,
therefore, to have far-reaching consequences for both fundamental
and applied research.

\begin{thebibliography}{10}
\bibitem{GIXD} Details of the GIXD experiment are available at
Science Online.
\bibitem{Future} Because of insufficient reproducibility, the
discussion of this pattern must be deferred.
\end{thebibliography}


\bibliographystyle{Science}

\begin{scilastnote}
\item This work was supported by the U.S. Department of Energy (DOE) grant
DE-FG02-88-ER45379 and the U.S.-Israel Binational Science
Foundation, Jerusalem. We acknowledge beamline assistance from J.
Gebhardt, T. Graber, and H. Brewer at the Chemistry and Materials
Science sector of the Center for Advanced Radiation Sources
(ChemMatCARS). ChemMatCARS Sector 15 is principally supported by
NSF/DOE grant CHE0087817. The Advanced Photon Source is supported
by the U.S. DOE contract W-31-109-Eng-38.
\\

4 April 2006; accepted 25 May 2006\\
10.1126/science.1128314
\end{scilastnote}



\clearpage

\clearpage
\begin{figure}
\vspace{-5mm} \centering
\includegraphics[angle=0,width=1.0\columnwidth]{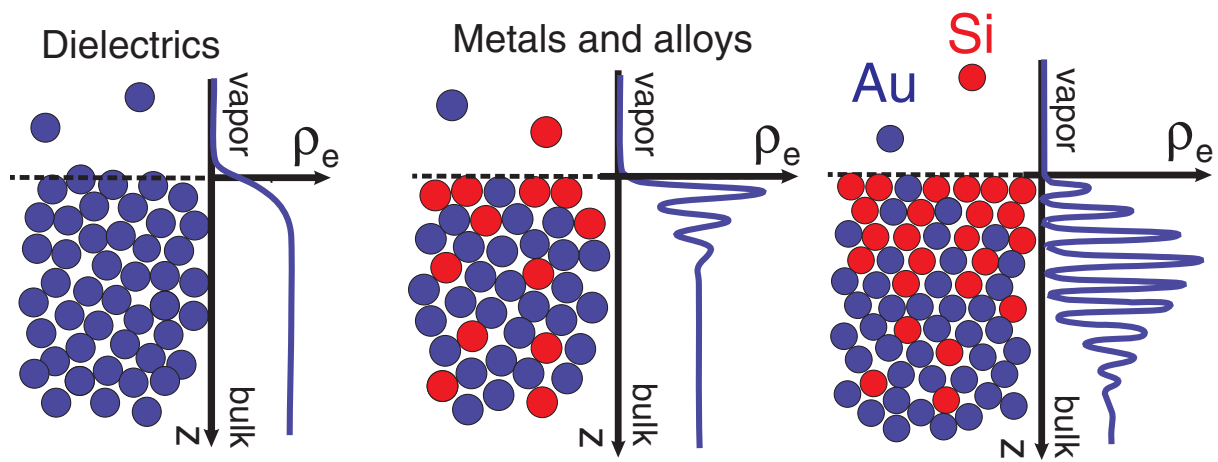}
\caption {{\bf Fig. 1.} Typical atomic surface structure and
corresponding electron density profiles $\rho_e(z)$ of nonlayered
dielectric liquids (left), standard layering in liquid metals and
alloys (middle) and enhanced layering in AuSi (right).}
\label{fig:ausi_fig1}
\end{figure}
\clearpage
\newpage

\begin{figure}
\vspace{-5mm} \centering
\includegraphics[angle=0,width=1.0\columnwidth]{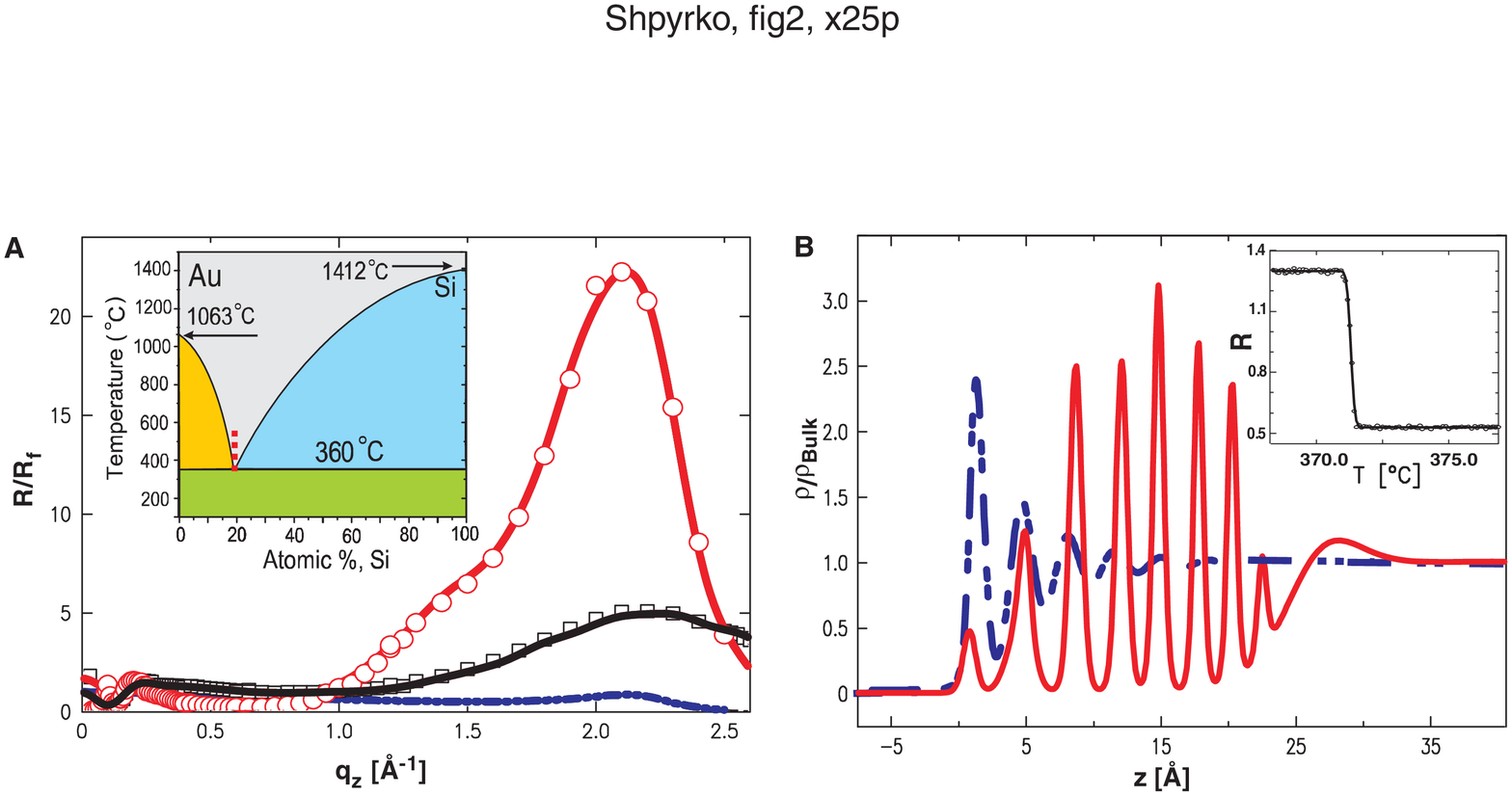}
\caption{{\bf Fig. 2.}  \textbf{(A)} Fresnel-normalized x-ray
reflectivity at 370~$^\circ$C (circles) and 375~$^\circ$C
(squares), the corresponding model fits (red and black lines,
respectively), and the curve expected for standard layering
(dashed blue line). \textbf{(Inset)} Bulk phase diagram of AuSi.
Grey area represent a liquid mixture phase, and the blue and
yellow areas indicate phase coexistence of solid Si or Au with a
liquid alloy. The green area corresponds to phase separated solid
Au and solid Si. \textbf{(B)} Surface-normal electron density
profiles corresponding to the same-line models in (A).
\textbf{(Inset)} Reflectivity at fixed $q_z=1.0$~{\AA}$^{-1}$
(circles) versus temperature, with a fit (black line) by an error
function centered at 371.29~$^\circ$C with a width of
0.17~$^\circ$C.} \label{fig:ausi_fig2}
\end{figure}
\clearpage
\newpage
\begin{figure}
\vspace{-0mm} \centering
\includegraphics[angle=0,width=0.85\columnwidth]{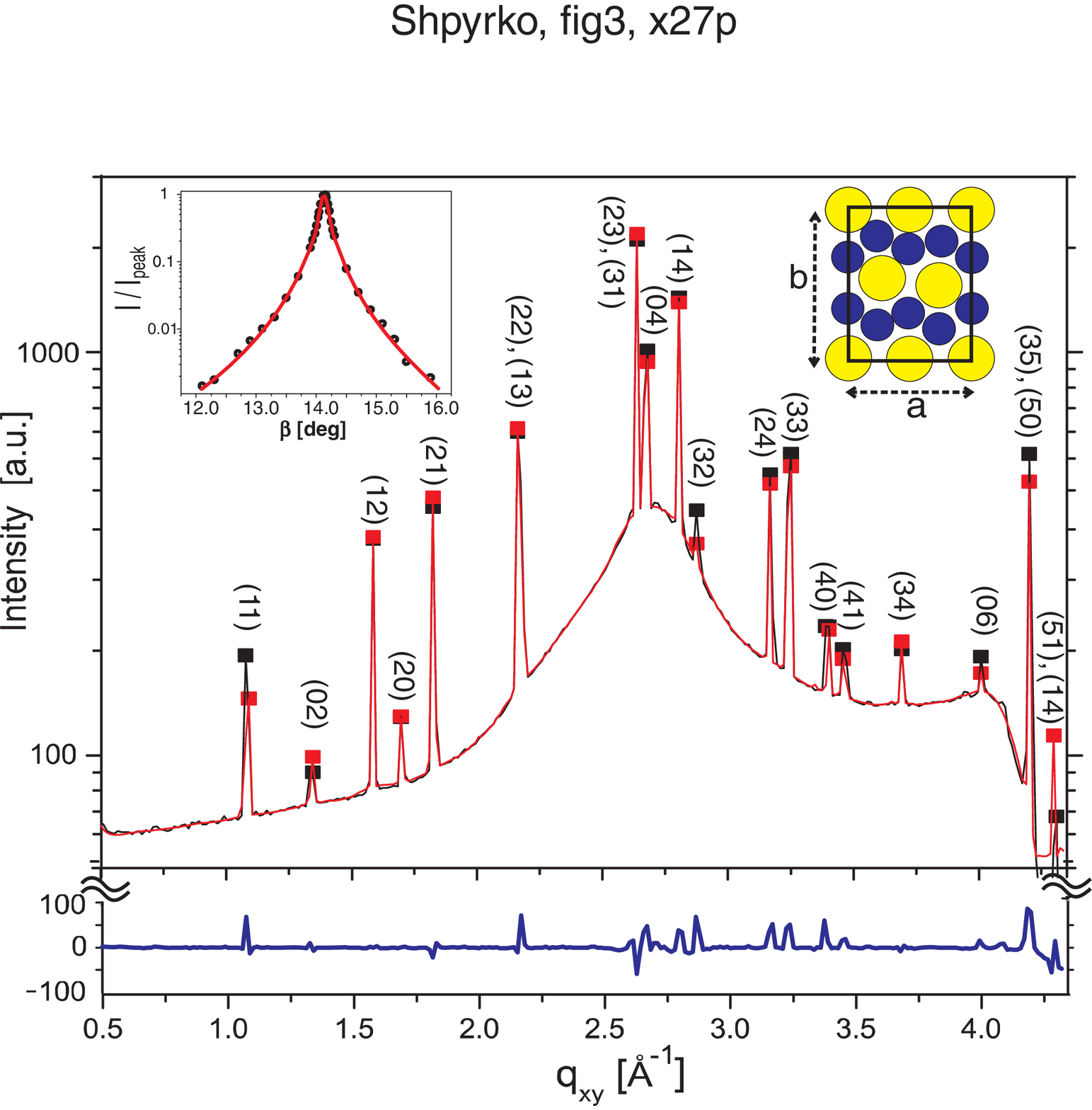}
\caption{{\bf Fig. 3.} Measured GIXD pattern (black line),
theoretical fit (red line), and their difference (blue line) for
the $359~^{\circ}\textrm{C}<T< 371~^\circ$C surface phase. a.u.,
arbitrary unites. (\textbf{Left inset)} Diffuse scattering profile
versus the output detector angle $\beta$ for a fixed incidence
angle $\alpha=14.11^{\circ}$ and its fit (line) by the capillary
wave theory prediction for $\gamma=780~$mN/m. $I/I_{peak}$ is the
x-ray intensity normalized by the scan's peak intensity value.
\textbf{(Right inset)} Crystal unit cell obtained from GIXD
pattern, where (\textbf{a}=7.386~{\AA}, \textbf{b}=9.386~{\AA}.
Au, yellow circles; Si, blue circles.} \label{fig:ausi_fig3}
\end{figure}

\clearpage

\end{document}